\documentclass[conference]{IEEEtran}
\ifCLASSINFOpdf
\else
\fi
\usepackage{amsmath, amsfonts, amssymb, graphicx, float}
\usepackage{amsbsy}
\usepackage{mathrsfs}
\usepackage{bm}
\usepackage{array,booktabs}
\usepackage{verbatim}
\usepackage{multirow}
\usepackage{fancyhdr}
\usepackage[T1]{fontenc}
\usepackage{textcomp}
\usepackage[scaled]{helvet}
\begin{document}
%
\title{Efficient network navigation with partial information}

\author{\IEEEauthorblockN{Xiaoran Yan}
\IEEEauthorblockA{Indiana University Network Science Institute (IUNI)\\
Indiana University, Bloomington, Indiana 47408\\
Email: xiaoran.a.yan@gmail.com}
\IEEEauthorblockN{Olaf Sporns}
\IEEEauthorblockA{Department of Psychological and Brain Sciences\\
\& Indiana University Network Science Institute (IUNI)\\
Indiana University, Bloomington, Indiana 47405}
\IEEEauthorblockN{Andrea Avena-Koenigsberger}
\IEEEauthorblockA{University Information Technology Services\\
Indiana University, Bloomington, Indiana 47405}
}

%


\maketitle


\begin{abstract}
  We propose a information theoretical framework to capture transition and information costs of network navigation models. Based on the minimum description length principle and the Markov decision process, we demonstrate that efficient global navigation can be achieved with only partial information. Additionally, we derived a scalable algorithm for optimal solutions under certain conditions. The proposed algorithm can be interpreted as a dynamical process on network, making it a useful tool for analysing and understanding navigation strategies on real world networks.
\end{abstract}



%
\IEEEpeerreviewmaketitle

\section{Introduction}
The function of many real world networks relies on physical or information exchange between their constituent elements. While efficient navigation or communication across large distances is widely observed in real world systems \cite{travers1967small}, the underlying mechanism is still not completely understood. Previous work on efficient navigation on networks highlighted the importance of small-world topologies for creating short-cuts at low wiring cost \cite{watts_collective_1998}. Yet, such a navigation model assumes that system's elements having information about the global topology and therefore shortest paths connecting the constituent elements  \cite{goni2013exploring}. A more detailed analysis of navigation strategies should not only concern about navigation efficiency, but also takes into account the cost of storing and using such topological information.

A drastically different picture emerges if we consider a navigation model that use minimal topological information, i.e. the unbiased random walk (URW) or the uniform diffusion process \cite{chung_heat_2007,lambiotte_laplacian_2008}. While highly efficient in terms of information cost, its transition cost is much higher in terms of number of steps to reach a specific destination. In this paper, we call the time cost of information routing as the \emph{transmission cost}, whereas the cost of storing and using topological information as the \emph{information cost}.

Between the two extremes of random walk and shortest path routing, a spectrum of communication processes deserve greater attention. For example, a preferential choice policy where nodes route messages towards high degree nodes can decrease search times significantly \cite{adamic2001search}, yet the information cost is small since nodes only need to know the degree of their neighbors. In general, biased random walks (BRW) can generate relatively efficient communication processes and are able to account for many navigation patterns observed in real world systems \cite{lambiotte_laplacian_2008,csimcsek2008navigating}. 

Here, we propose a information theoretical framework based on BRW to capture transition and information costs in terms of minimum description coding length. Optimization of the BRW can be elegantly solved as a Markov decision process (MDP) \cite{BellmanMDP, todorov_efficient_2009}. Under certain conditions, the optimal solution of the MDP reduces to a linear eigenvalue problem. This new algorithm is also theoretically connected with the dynamical process framework called Z-Laplacian \cite{yan_graph_2017}, leading to intuitive interpretations of the navigation strategies on real world networks. The proposed framework unifies a family of navigation strategies that include shortest path routing and random walk processes as special cases. The trade-off between information and transition costs under this framework demonstrates that efficient global navigation is possible with limited information.

\section{A navigation model with information cost}
Based on information coding theory, we first develop a unifying network navigation framework to quantify both transition and information costs. 
\subsection{The BRW model}
We consider a weighted and undirected graph $G = (V,E,\mathbb{A})$ composed of the size $n$ vertex set $V$ and the size $m$ edge set $E$. The graph topology is described by the symmetric non-negative matrix $\mathbb{A}$, where the generic element $\mathbb{A}_{uv} = \mathbb{A}_{vu} > 0$ if there is a weighted edge between vertices $u$ and $v$, while $\mathbb{A}_{uv} = \mathbb{A}_{vu} = 0$, otherwise. 

The simple discrete time URW on graph $G$, can be represented using a $n\times n$ stochastic matrix $\mathbb{P}_0 = \mathbb{A} \mathbb{D}^{-1}$, where $\mathbb{D}$ is the diagonal out-degree matrix, with entries $\mathbb{D}_u = \sum_v \mathbb{A}_{uv}$. At the other extreme, we have the shortest path routing. We capture the pair-wise distances between vertex pairs using the matrix $\mathbb{S}$. In particular, between the immediate neighbors $(u,v)\in E$, the shortest path distances are defined by inverting the non-zero entries of $\mathbb{A}$, as $\mathbb{S}_{uv} = 1/\mathbb{A}_{uv}$. For all other vertex pairs $(u,v)\notin E$, we solve the shortest path problem based on these distances\footnote{One of the earliest shortest path algorithms, the Bellman-Ford algorithm, is also proposed by Richard Bellman\cite{bellman1958routing}}.

To unify these two extreme cases and other more general navigation strategies, we define a family of target specific BRW with the probability functions $P(v|u,t)$, parameterized by the neighboring vertex $v$, conditioned on the source vertex $u$ and the target vertex $t$. The two extreme cases can be casted as special cases as follows:
\begin{align}
\label{eq:baseline}
P_0(v|u,t) =& P_0(v|u) = \frac{\mathbb{A}_{uv}}{\mathbb{D}_u}, \forall t \nonumber\\
P_{\text{SP}}(v|u,t) =& \begin{cases}
			  \frac{1}{|\mathbb{S}^v_{ut}|} & \parbox[t]{.4\textwidth}{if $\mathbb{S}_{ut} = \mathbb{S}_{uv}+\mathbb{S}_{vt}$,}\\
			  0 & otherwise,
			\end{cases}
\end{align}
where $|\mathbb{S}^v_{ut}|$ represents the number of unique neighbors $v$ that lie in one of the shortest paths between $u$ and $t$.

\subsection{The continuous action space MDP}
To cast the network navigation problem as a MDP with continuous action space \cite{russell2016artificial,todorov_efficient_2009}, we formally specify its five defining elements as $\{V, u_0, T, \mathcal{A}, l(u,T,\mathcal{A}) \}$, where
\begin{enumerate}
 \item $V$ is the state space, and we denote the current state, i.e. the position of the random walker, as $u\in V$, and the next state $v\in V$;
 \item $u_0\in V$ is the starting state;
 \item $T\subseteq V$ is the set of potential target/absorbing states;
 \item A continuous action space $\mathcal{A}$, which is directly represented by the transition probability function $P_\mathcal{A}(v|u,t)$;
 \item $l(u,t,\mathcal{A})$ is the immediate loss (or ``negative reward'') function by taking action $\mathcal{A}$ at state $u$ with the target $t\in T$, which leads to the transition probabilities $P_\mathcal{A}(v|u,t)$. Mathematically, it is defined as $V\times V \times \mathcal{A}\rightarrow \mathbb{R}_+$. 
\end{enumerate}

Based on the above definition, we can formulate a Bellman equation, which is fundamental for solving MDP, optimal control and dynamic programming problems in general \cite{BellmanDP}.
\begin{equation}
\label{eq:Bellman}
 L(u,t) = \min_{\mathcal{A}}\left\{l(u,t,\mathcal{A})+E_{v\sim P_\mathcal{A}(v|u,t)}[L(v,t)]\right\}\;,
\end{equation}
where the optimal loss $L(u,t)$ is the minimum (over possible actions, or transition probabilities) of the immediate loss $l(u,t,\mathcal{A})$ plus the expected cost-to-go at the next state $v$. The expected cost-to-go function can take several different formulations, in this paper we focus on the undiscounted infinite-horizon total-cost problems with absorbing states, which follows the above recursive formulation.

To capture both transition and information costs, we further define the immediate loss function as,
\begin{equation}
\label{eq:totalCost}
 l(u,t,\mathcal{A}) = \gamma C^{\text{trans}}(u) + C^{\text{info}}(u,t,\mathcal{A})\;,
\end{equation}
where $C^\text{{info}}$ represents the information cost needed for taking the action $\mathcal{A}$. In optimal control theory, $C^\text{{info}}$ can also be understood as the control cost. $C^{\text{trans}}$ is the transition cost of the BRW step from vertex $u$, and we set it to zero for target/absorbing states \footnote{The MDP framework allows for arbitrary choice of $C^{\text{trans}}$, as we will show in our derivations. In our experiments,  however, we generally set it to be 1 to represent random walk hops.}. $\gamma$ is a tuning factor for balancing between the transition and information costs. A bigger $\gamma$ will increase the penalty of transition cost in the overall MDP optimization.

\subsection{Defining the information cost}
To quantify the information cost, we start with an explicit routing model from the computer networking literature \cite{peleg1989routing}. We first consider the target agnostic URW, which can be represented by its transition probabilities $P_0(v|u)$. Therefore, each step at vertex $u$ reads from a routing table of size $\mathcal{O}(|\delta(u)|*\log n)$, where $\delta(u)$ is the set of the immediate neighbors of $u$, and $\log n$ is the binary bits needed to index each vertex. We also assume edge weights/distances and probability densities can be encoded in constant length.  

Adapting the explicit routing model for our BRW model, each vertex $u$ needs a routing table to encode its transition probability function $P_\mathcal{A}(v|u,t)$. Therefore, each step at vertex $u$ reads from a routing table of size $\mathcal{O}(|\delta(u)|*(|T|+1)*\log n)$, where $|T|$ is size of the target set. Notice here a independent transition probability function $P_\mathcal{A}(v|u,t)$ is fully specified for each $t\in T$, leading to a $\mathcal{O}(k)$ fold cost increase when compared with the baseline $P_0$. The original $P_0$ table is also kept for routing flexibility. 

The aforementioned explicit routing table size is an example of intuitive but inefficient information representation. This is especially the case when the BRW process only deviates the slightest from the URW: it still incurs the full $\mathcal{O}(k)$ fold cost increase despite the fact that only minimal routing information is utilized. According to coding theory \cite{shannon1949mathematical}, we can save on information cost by compressing the topological information with more efficient coding schemes \cite{mackay_2003}. In fact, there exists optimal coding schemes that can approach the theoretical limit of a given BRW model, which is called the minimum description length (MDL) \cite{grunwald2007}.

For better intuitions and mathematical properties essential for later derivations, we choose to adopt the ``crude'' two-part code for measureing the MDL \cite{grunwald2007}. In two-part codes, the routing table is only encoded once, and all random walk steps will be encoded using the optimal code based on the table. In this way, we can encode a BRW step from the vertex $u$ using the reference transition probabilities $P_0$,
\begin{align}
\label{eq:MDL}
\text{MDL}(\mathcal{A}|u, t, P_0) = \beta\text{MDL}_1[P_0] + \text{MDL}_2[\mathcal{A}|u, t, P_0] \;.
\end{align}
The first part $\text{MDL}_1$ captures the MDL needed to encode the reference URW transition matrix $P_0$. We used the fractional constant $\beta$ to distribute the cost of this one-time code across all BRW steps. Based on our explicit routing model analysis, we already have $\text{MDL}_1[P_0] = |\delta(u)|*\log n$, where the $\mathcal{O}$ notation is also absorbed into the constant $\beta$.

The second part $\text{MDL}_2$, or the minimal number of bits required to encode an outcome of the BRW distribution $P_\mathcal{A}$ using the reference model $P_0$, is captured by the following cross entropy,
\begin{align*}
\text{MDL}_2[\mathcal{A}|u, t, P_0] =  -\sum_{v \in \delta(u)} P_\mathcal{A}(v|u,t)\log P_0(v|u)\;.
\end{align*}
Putting them together, we have,
\begin{align*}
  &\text{MDL}(\mathcal{A}|u, t, P_0)\\
 =& \beta |\delta(u)|\log n -\sum_{v \in \delta(u)} P_\mathcal{A}(v|u,t)\log P_0(v|u)\;.
\end{align*}

Similarly we can quantify the MDL of a BRW step using the full BRW routing table,
\begin{align*}
  &\text{MDL}(\mathcal{A}|u, t, P_\mathcal{A})\\
 =& (|T|+1)\beta|\delta(u)|\log n -\sum_{v \in \delta(u)} P_\mathcal{A}(v|u,t)\log P_\mathcal{A}(v|u,t)\;.
\end{align*}

To better compare the MDLs under the MDP model, we compare their cumulative difference over the random walk trajectory $\varGamma$, 
\begin{align}
\label{eq:cumulativeInfo}
  &\text{MDL}(\varGamma|u, t, P_0) - \text{MDL}(\varGamma|u, t, P_\mathcal{A})  = -|T||\mathbb{D}|\log n \nonumber\\
  & -\sum_{u \in \varGamma} \sum_{v \in \delta(u)} P_\mathcal{A}(v|u,t)\log \frac{P_0(v|u)}{P_\mathcal{A}(v|u,t)}  \nonumber\\
 =& -|T||\mathbb{D}|\log n + \sum_{u \in \varGamma}  D_{\text{KL}}(P_\mathcal{A}(v|u,t)||P_0(v|u)) \;,
\end{align}
where $|\mathbb{D}|$ represents the total degree of the network.

If \eqref{eq:cumulativeInfo} is greater than $0$, we can always save information cost by paying for the full routing table up front. In this case, the MDP optimization in \eqref{eq:totalCost} becomes trivial with a constant information cost. No matter how we define the transition cost $C^{\text{trans}}(u)$ or the parameter $\gamma$, the total coding length is guaranteed to be shorter by taking the shortest path instead of the random walk trajectory $\varGamma$. Because $\varGamma_{\text{SP}} \in \mathbb{S}$ is by definition optimal in transition costs. 

When \eqref{eq:cumulativeInfo} is less than $0$, however, it can potentially be more efficient by using the simpler routing model of URW as the reference. We only pay for the additional information to manipulate the transition probabilities as the BRW proceeds, captured by the KL-divergence in \eqref{eq:cumulativeInfo}. In this case, the MDP becomes a trade-off between information and transition costs. To minimize the additional cost, we define the KL-divergence as the information cost of the immediate loss function, 
\begin{align}
\label{eq:immediateInfo}
  C^\text{{info}}(u,t,\mathcal{A}|P_0) = D_{\text{KL}}(P_\mathcal{A}(v|u,t)||P_0(v|u)) \;.
\end{align}

While the optimization focuses on \eqref{eq:immediateInfo}, we need to make sure the first term in \eqref{eq:cumulativeInfo} is greater than the cumulative KL-divergence to avoid the trivial solution. Here the first term $|T||\mathbb{D}|\log n$ can also be interpreted as the additional information cost for the full routing tables. To simplify notations, we shall consider the target $t$ as given for the rest of the paper.


\subsection{Solving the Bellman equation}
With information cost properly defined, we put it together with transition cost to complete the definition of the immediate loss function.
\begin{align}
	\label{eq:BRWcosts}
	&l(u,\mathcal{A}) =\gamma C^{\text{trans}}(u)  + C^{\text{info}}(u,\mathcal{A}|P_0) \nonumber\\
                 =& \gamma C^{\text{trans}}(u) + \sum_{v \in \delta(u)} P_\mathcal{A}(v|u)\log \frac{P_\mathcal{A}(v|u)}{P_0(v|u)}\;,
\end{align}

Traditional MDP solvers such as value iteration and policy iteration algorithms require multiple convergence steps and quickly become intratable in large networks. In their seminal work \cite{todorov_linearly-solvable_2007,todorov_efficient_2009}, the authors developed an efficient linear solver for MDPs when the information cost follows the KL-divergence. This is exactly the formulation of our MDP model in \eqref{eq:BRWcosts}. Substitute the immediate loss function to the Bellman equation \eqref{eq:Bellman},
\begin{align}
\label{eq:Bellman2}
&L(u) = \min_{\mathcal{A}}\left\{l(u,\mathcal{A})+E_{v\sim P_\mathcal{A}(v|u)}[L(v)]\right\}\nonumber\\
=&\gamma C^{\text{trans}}(u) + \min_{\mathcal{A}}\left\{ E_{v\sim P_\mathcal{A}(v|u)} \left[\log \frac{P_\mathcal{A}(v|u)}{P_0(v|u)} + L(v)  \right] \right\} \nonumber\\
=&\gamma C^{\text{trans}}(u) + \min_{\mathcal{A}}\left\{\sum_{v\in \delta(u)} P_\mathcal{A}(v|u) \left[\log \frac{P_\mathcal{A}(v|u)e^{L(v)}}{P_0(v|u)}\right]\right\}\nonumber\\
=&\gamma C^{\text{trans}}(u) +\nonumber\\
 &\min_{\mathcal{A}}\left\{\sum_{v\in \delta(u)} \frac{P_0(v|u) e^{\alpha_v}}{Z_\mathcal{A}(u)}  \left[\log \frac{P_0(v|u) e^{\alpha_v} e^{L(v)}}{Z_\mathcal{A}(u)P_0(v|u)}\right]\right\}\;.
\end{align}
Here we have rewrite the transition probability of the BRW in terms of the reference URW 
\begin{align}
\label{eq:reformulation}
P_\mathcal{A}(v|u) = \frac{P_0(v|u) e^{\alpha_v}}{Z_\mathcal{A}(u)}\;,
\end{align}
where $Z_\mathcal{A}(u) = \sum_{v\in \delta(u)} P_0(v|u) e^{\alpha_v}$ is the normalizing constant such that $\sum_{v\in \delta(u)} P_\mathcal{A}(v|u) = 1$. This reformulation allows us to rescale $P_\mathcal{A}$ to arbitrary distributions, with only one limitation that $P_\mathcal{A}(v|u)$ will remain at zero if $P_0(v|u)=0$.

The minimization of \eqref{eq:Bellman2} can be performed in closed form using Lagrange multipliers, as follows. For each state $u$, we define the Lagrangian as follows,
\begin{align}
\label{eq:Lagrangian}
&\mathcal{L}(\mathcal{A}, \lambda_u) = \sum_{v\in \delta(u)} \frac{P_0(v|u) e^{\alpha_v}}{Z_\mathcal{A}(u)} \log \frac{e^{(\alpha_v+L(v))}}{Z_\mathcal{A}(u)} \nonumber\\  
                 &- \lambda_u  \left[ \sum_{v\in \delta(u)}\frac{P_0(v|u) e^{\alpha_v}}{Z_\mathcal{A}(u)} - 1\right] \nonumber\\  
                =& \sum_{v\in \delta(u)} \frac{P_0(v|u)e^{\alpha_v}}{Z_\mathcal{A}(u)} \left[\alpha_v+L(v)-\log Z_\mathcal{A}(u) \right]\nonumber\\  
                 &- \lambda_u  \left[ \sum_{v\in \delta(u)}\frac{P_0(v|u) e^{\alpha_v}}{Z_\mathcal{A}(u)} - 1\right] 
\end{align}

Taking partial derivative with respect to $\alpha_v$, we have
$$\frac{\partial\mathcal{L}}{\partial \alpha_v} = \frac{P_0(v|u)}{Z_\mathcal{A}(u)}  e^{\alpha_v} (\alpha_v+L(v)-\log Z_\mathcal{A}(u) +1-\lambda_u)
$$

Setting $\frac{\partial\mathcal{L}}{\partial \alpha_v} = 0$, with $P_0(v|u)>0$, we get
\begin{align}
\label{eq:optimal}
\alpha_v^*  = -L(v)+\log Z_\mathcal{A}(u)  -1 + \lambda_u\;.
\end{align}

Checking the second derivative at the optimal value $\alpha_v^*$,
$$\left.\frac{\partial^2\mathcal{L}}{(\partial \alpha_v)^2}\right|_{\alpha_v=\alpha_v^*} = \frac{P_0(v|u)}{Z_\mathcal{A}(u)}  e^{\alpha_v^*} >0\;,
$$
which confirms $\alpha_v^*$ is a minimizer of \eqref{eq:Bellman2}. Combine it with \eqref{eq:reformulation} with the normalization constrain, we have
\begin{align}
\sum_{v\in \delta(u)} P^*_\mathcal{A}(v|u) =& \sum_{v\in \delta(u)} \frac{P_0(v|u) e^{-L(v)+\log Z_\mathcal{A}(u)  -1 + \lambda_u}}{Z_\mathcal{A}(u)}\nonumber\\  
                                        1  =& e^{\lambda_u-1} \sum_{v\in \delta(u)} P_0(v|u) e^{-L(v)}\nonumber\\  
                                \lambda_u  =& -\log \left[ \sum_{v\in \delta(u)} P_0(v|u) e^{-L(v)} \right] +1\;.
\end{align}

Plug it back to \eqref{eq:optimal}, we have the close form solution of the optimal action,
\begin{align}
\label{eq:optimal2}
\alpha_v^* =& -L(v)+\log Z_\mathcal{A}(u) -\log \left[ \sum_{v\in \delta(u)} P_0(v|u) e^{-L(v)} \right]\nonumber\\  
           =& -L(v)+\log\frac{Z_\mathcal{A}(u)}{Z_L(u)} - \log\sum_{v\in \delta(u)} \frac{P_0(v|u) e^{-L(v)}}{Z_L(u)} \nonumber\\  
           =& -L(v)+\log Z_\mathcal{A}(u)-\log Z_L(u) \;,
\end{align}
where $Z_L(u) = \sum_{v\in \delta(u)} P_0(v|u) e^{-L(v)}$ is the normalizing constant such that summation in the third term of \eqref{eq:optimal2} is over an proper distribution. Finally, by plugging it back to \eqref{eq:reformulation}, we can rewrite the optimal transition probabilities as
\begin{align}
\label{eq:reformulation2}
P^*_\mathcal{A}(v|u) = \frac{P_0(v|u) e^{\alpha_v^*}}{Z_\mathcal{A}(u)} = \frac{P_0(v|u) e^{-L(v)}}{Z_L(u)}\;.
\end{align}

This reformulated optimal transition probabilities \eqref{eq:reformulation2} has an important consequence. Its only dependence on the current state is through the normalizing constant $Z_L(u)$. The real deciding factor is the cost-to-go at the next state $L(v)$. A even more intuitive interpretation emerges if we put it back into \eqref{eq:Bellman2},
\begin{align}
\label{eq:Bellman3}
&L(u) = \min_{\mathcal{A}}\left\{l(u,\mathcal{A})+E_{v\sim P_\mathcal{A}(v|u)}[L(v)]\right\}\nonumber\\
=&\gamma C^{\text{trans}}(u) + \min_{\mathcal{A}}\left\{\sum_{v\in \delta(u)} P_\mathcal{A}(v|u) \left[\log \frac{P_\mathcal{A}(v|u)e^{L(v)}}{P_0(v|u)}\right]\right\}\nonumber\\
=&\gamma C^{\text{trans}}(u) + \sum_{v\in \delta(u)} \frac{P_0(v|u) e^{-L(v)}}{Z_L(u)} \left[\log \frac{1}{Z_L(u)}\right]\nonumber\\
=&\gamma C^{\text{trans}}(u) - \log[Z_L(u)] + D_{\text{KL}}(P^*_\mathcal{A}(v|u)||P^*_\mathcal{A}(v|u))\nonumber\\
=&\gamma C^{\text{trans}}(u) - \log\sum_{v\in \delta(u)} P_0(v|u) e^{-L(v)}\;.
\end{align}

Here, the optimal cost $L(u) = \gamma C^{\text{trans}}(u) - \log[Z_L(u)]$ is achieved when the reformulated KL-divergence is zero. This is exactly the intuition of the optimal action in \eqref{eq:reformulation2}, where the reference transition probabilities are scaled according to their cost-to-go in the future state $L(v)$. To solve for the optimal cost function recursively, we exponentiate both sides of \eqref{eq:Bellman3},
\begin{equation}
\label{eq:exploss}
\exp[-L(u)]  = \exp[-\gamma C^{\text{trans}}(u)] \sum_{v\in \delta(u)} P_0(v|u) \exp[-L(v)]\;.
\end{equation}

In matrix notations, this is equivalent to 
\begin{equation}
\label{eq:eigen}
 \overrightarrow{e} = \mathbb{T} \mathbb{P}_0 \overrightarrow{e} \;,
\end{equation}
where the vector component $\overrightarrow{e_u}$ represents the exponentiated loss $\exp[-L(u)] $, the matrix $\mathbb{T}$ is the diagonal matrix with entries $ \exp[-\gamma C^{\text{trans}}(v)]$, and $\mathbb{P}_0$ is the reference stochastic matrices. We have now successfully transformed our MDP into a linear eigenvalue problem. \footnote{This can be understood as a special case of \cite{todorov_linearly-solvable_2007,todorov_efficient_2009}. Keep in mind here we have suppressed the notation for $t$, but all matrices are target dependent.}

The solution, which is the largest eigenvalue of $\mathbb{T} \mathbb{P}_0$, will in turn determine the optimal transition probabilities,
\begin{equation}
\label{eq:optimalA}
 \mathbb{P}^*_\mathcal{A} \propto  \mathbb{E}  \mathbb{P}_0 \;,
\end{equation}
where the diagonal matrix $\mathbb{E}$ is composed of the components of vector $\overrightarrow{e}$. 

Plug this back into \eqref{eq:cumulativeInfo}, we can bound the cumulative information cost over the trajectory $\varGamma$ of the optimal BRW,
\begin{align}
  \label{eq:infoCostT}
 &\sum_{u \in \varGamma} D_{\text{KL}}(P^*_\mathcal{A}(v|u)||P_0(v|u)) \nonumber\\
 \le& |\varGamma| \max_{u}[D_{\text{KL}}(P^*_\mathcal{A}(v|u)||P_0(v|u))] \;,
\end{align}
where we adopted a simple bound for model analysis. When \eqref{eq:infoCostT} is smaller than $|T||\mathbb{D}|\log n$, i.e., the cumulative KL-divergence is cheaper to encode than the full routing tables, satisfying the condition that \eqref{eq:cumulativeInfo} is smaller than zero. The solution in \eqref{eq:optimalA} is therefore also globally optimal. 

\section{Model analysis}
With the navigation model properly defined and its optimization problem solved, we further investigate its properties. 
\subsection{Random walk and shortest path navigation}
We start with a theoretical derivations of the two extreme cases, i.e. $P_0$ and $P_{\text{SP}}$ in \eqref{eq:baseline}. The key parameter here is $\gamma$, which controls the trade-off between the transition and information costs. Recall the immediate cost is of the form,
\begin{align*}
	\label{eq:BRWcosts}
	&l(u,\mathcal{A}) =\gamma C^{\text{trans}}(u)  + C^{\text{info}}(u,\mathcal{A}|P_0) \nonumber\\
                 =& \gamma C^{\text{trans}}(u) + \sum_{v \in \delta(u)} P_\mathcal{A}(v|u)\log \frac{P_\mathcal{A}(v|u)}{P_0(v|u)}\;.
\end{align*}

To recover the random walk model $P_0$, we simply set $\gamma=0$. Intuitively, with no penalty to transition cost, the MDP optimization focuses on minimizing the information cost, which is achieved when the KL-divergence between $P_\mathcal{A}$ and $P_0$ is zero, or $P^*(v|u,t) = P_0(v|u)$. Combined with the fact that in our MDL model, the routing table of $P_0$ is also more efficient to encode, the BRW indeed reduces to the reference $P_0$

This is further confirmed by the optimal total cost in \eqref{eq:eigen}. Matrix $\mathbb{T}$ becomes the identity matrix when $\gamma=0$, leading to the classic eigenvalue problem of the random walk $P_0$, 
$$ \overrightarrow{e} = \mathbb{P}_0 \overrightarrow{e} \;.
$$

Following the same logic, we set $\gamma=\infty$. From the perspective of the immediate cost, the MDP optimization will now focus on minimizing the transition cost, which is achieved when the shortest paths are taken. 

However, things are less clear from the perspective of the optimal total cost in \eqref{eq:eigen}. Mathematically, as $\gamma$ goes to infinity, all entries of $\mathbb{T}$ becomes vanishingly small. The only exception is the target state $t$, which by our definition has $C^{\text{trans}}(t) = 0$. The resulting matrix product $\mathbb{T} \mathbb{P}_0$ is therefore dominated by the row corresponding to the target state $t$, funneling BRW trajectories towards the shorter paths.

To better understand the effect of the parameter $\gamma$ and the matrix product $\mathbb{T} \mathbb{P}_0$, we will re-investigate the mathematical optimization in \eqref{eq:eigen}. One numerical solution to this eigenvalue problem is to use iterative methods as follows,
\begin{equation}
\label{eq:interative}
 \overrightarrow{e}_{t+1} = \mathbb{T} \mathbb{P}_0 \overrightarrow{e}_t \;.
\end{equation}

If we treat the vectors $\overrightarrow{e}_t$ as states of the nodes and $\mathbb{T} \mathbb{P}_0$ as a graph operator, \eqref{eq:interative} becomes a special dynamical process under the Z-Laplacian framework \cite{yan_graph_2017}.

The Z-Laplacian framework provides an intuitive interpretation for \eqref{eq:interative} as a shrinking dynamical process. $\mathbb{T}$ represents the shrinking replicating factor, where the incoming random walk traffic to $v$ through $\mathbb{P}_0$ is discounted by the factors $\exp[-\gamma C^{\text{trans}}(v)]$. This discounting effect is compounded by local network structure on top of these shrinking factors. The overall diffusion gets weaker as the process moves further from $t$, funneling trajectories towards the target.

\subsection{Efficient navigation on real world networks}
To verify the mathematical intuitions, we will empirically investigate the behavior of our model on real world networks. We first numerically solve for the eigen value problem of \eqref{eq:eigen}, and recover the optimal transition probabilities as in \eqref{eq:optimalA}. We then compare the navigation performance at several $\gamma$ values, with the full routing table model as the baseline. For both URW and BRW models, we will use mean first passage time \cite{grinstead2012introduction} to measure the "number of hops" as the transition cost. This is consistent with the transition cost in \eqref{eq:totalCost}, when we set $C^{\text{trans}}(u) = 1$ for non-target nodes. The goal in this subsection is to demonstrate the feasibility and efficiency of global navigation without all the routing tables.

\begin{figure}
  \centering  
  \includegraphics[width=0.35\textwidth]{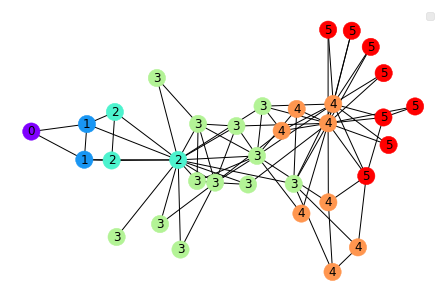}
  \includegraphics[width=0.35\textwidth]{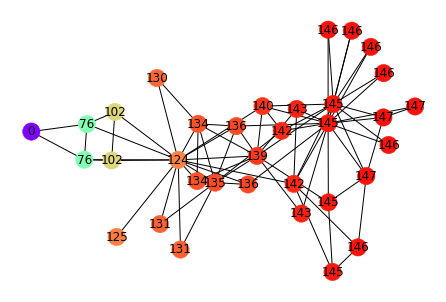}
  \caption{Shortest paths (top) and mean first passage time (bottom) to the leftmost node in Zachary's Karate Club. The color and integer label of each node represents the transition cost, or the number of hops (rounded for mean first passage time) requred to reach the leftmost node.}
  \label{fig:karate}
\end{figure}

The first network we study is a social network consisting of $34$ members of a karate club in a university, where undirected edges represent friendships \cite{ZacharyKarateClub}. This well-studied toy network is made up of two assortative blocks centered around the instructor and the club president, each with a high degree hub and lower-degree peripheral vertices. A visualization of the network is shown in Fig. \ref{fig:karate}.

Here we focus on the routing problem to the leftmost node or "node 16". As Fig. \ref{fig:karate} demonstrates, shortest path routing is much more efficient in terms of transition cost. However, it requires us to pay up front for the information cost of the routing tables, at $|\mathbb{D}|\log n = 793.6$ bits.

\begin{figure}
  \centering  
  \includegraphics[width=0.35\textwidth]{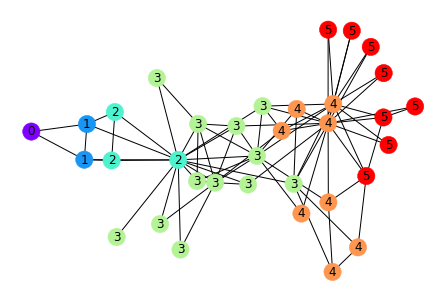}
  \includegraphics[width=0.35\textwidth]{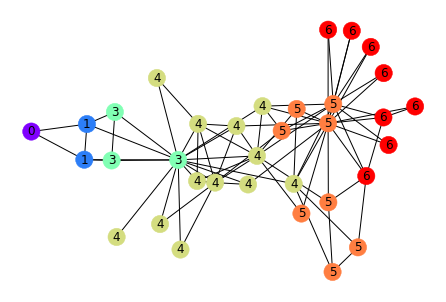}
  \caption{Mean first passage time to the leftmost node in Zachary's Karate Club under our optimal BRW models with  $\gamma = 50$ (top) and $\gamma = 2.5$ (bottom). The color and integer label of each node represents the transition costs, or the rounded mean first passage time to reach the leftmost node.}
  \label{fig:karate2}
\end{figure}

To explore the trade-offs between transition and information costs, we constructed two optimal BRW models (see Fig. \ref{fig:karate2}). With $\gamma = 50$, the optimal solution leads to mean first passage times that closely matches the shortest path hops. In this case, the BWR paid for the information to manipulate the transition probabilities from the reference $P_0$. According to \eqref{eq:infoCostT}, the upper bound for the information cost is $15.0$ bits \footnote{This upper bound is calculated by multiplying the largest "hop number" in the BWR (rounded for mean first passage time) with \eqref{eq:infoCostT}}, making it much more efficient than the full routing table model, under the setting of a undiscounted infinite-horizon total-cost problem with absorbing states. 

A more interesting solution emerges when we set $\gamma = 2.5$, leading to a better balance between transition cost and information cost. The former, measured in mean first passage time, becomes slightly higher. However, the cumulative information cost is now bounded by $13.0$ bits, making it a even more efficient solution overall, despite the increase in transition cost.

To further validate that efficient navigation can be achieved with partial information, we consider an airline network constructed from public data on flights between major airports around the world. The edges here are weighted representing the number of flights as well as the capacity of the plane model operating each flight \cite{openFlights}. This network contains 3253 airports/nodes, and 18997 direct flights/edges. Direct flights are treated as undirected edges because most are symmetric.

\begin{figure}
  \centering  
  \includegraphics[width=0.156\textwidth]{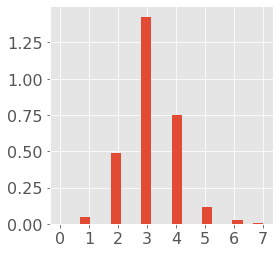}
  \includegraphics[width=0.156\textwidth]{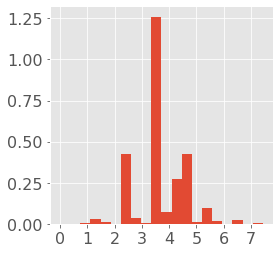}
  \includegraphics[width=0.156\textwidth]{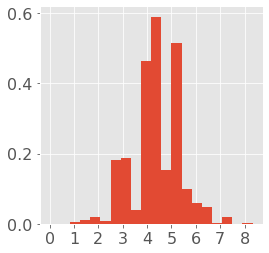}
  \caption{Shortest paths (left) and mean first passage time (middle $\gamma = 50$, right $\gamma = 5$) to LAX in the airline network. The histogram here captures the distribution of the number of hops (rounded to 20 bins for mean first passage time) requred to connect LAX with direct flights around the world.}
  \label{fig:flights}
\end{figure}

We focus on the routing problem to the Los Angeles International Airport (LAX) in this network. In Fig. \ref{fig:flights}, the middle panel is the mean first passage time of our optimal BRW models with $\gamma = 50$. It is able to approach shortest path routing (left panel) in terms of transition cost, while incurring a cumulateive information cost bounded by $56.1$ bits. Encoding all the shortest path routing tables will cost us $443297$ bits. The right panel with $\gamma = 5$ achieves even better efficiency at slightly higher transition cost, only paying up to $38.5$ bits in information cost over potential navigation trajectories.

Finally, we can verify that \eqref{eq:cumulativeInfo} is less than zero for all panels of Fig. \ref{fig:karate2} and Fig. \ref{fig:flights}, guaranteeing them as global optima under their respective $\gamma$ values.

\section{Conclusion and Future work}
In this paper, we proposed a information theoretical framework to capture transition and information costs of network navigation models. By casting it as a Markov decision process, we derived a scalable algorithm for optimal solutions. Empirically, we demonstrated that efficient global navigation can be achieved on real world networks with only partial information.

In the future, we plan to investigate the optimization problem of the $\gamma$ parameter under different horizon and cost formulations. More complex routing tasks where target nodes changes over the course of dynamical process is another direction. A deeper theoretical understanding under the Z-Laplacian framework will help to shred light on how such network control problems can be transformed into the graph diffusion framework. Potential applications include information communication patterns in the brain connectome and traffic routing optimization on road networks.
 
\bibliographystyle{IEEEtran}
\bibliography{sample-bibliography} 

\end{document}